\documentclass{kluwer}    
\begin{document}
\begin{opening} 

\title{Chemical Evolution of Bulges}
\author{Mercedes \surname{Moll\'{a}}}
\institute{Departamento de F\'{\i}sica Te\'{o}rica, C-XI,
  Universidad Aut\'{o}noma de Madrid, 28049 Madrid, Spain}
\author{Federico \surname{Ferrini}}
\institute{INTAS, 58 Avenue des Arts, 
1000 Bruxelles, Belgium}    

\runningauthor{M. Moll\'{a} \& F. Ferrini}
\runningtitle{Chemical evolution of Bulges}


\end{opening} 

\section{The Multiphase Model applied to Bulges}

We present the multiphase model applied to a set of bulges.  This
model was first applied to the Solar Neighborhood, and then to the
Galactic Disk, by computing the radial dependence of input parameters,
which govern the gas accumulation in the disk, and the cloud and star
formation processes. These processes result enhanced in the central
region due to the volume effect. The evolution of the Galactic bulge
was thus directly obtained \cite{MF}, with the same set of
efficiencies and characteristic collapse time scale defined for the
disk.  We have also applied the model to a sample of spiral disks, by
changing the input parameters according their total masses and Hubble
types.  With the same approach used for our Bulge, we now extend the
model to their bulges \cite{Molla}

The resulting SFR in the central regions shows a intense initial
episode in the first Gyr. Surface densities for atomic and molecular
gas of later type bulges are higher than those of earlier types.  The
predicted Galactic bulge mean metallicity, $\overline {\rm
[Fe/H]}=-0.17$, and the corresponding metallicity distribution, are in
excellent agreement with data. Mean stellar abundances are subsolar
for all modelled bulges ($\rm T \ge 3$), independently of the Hubble
type, the arm class and/or the luminosity of their host galaxies,
reproducing the estimations from color data. The ratio between past
and present SFR and the abundance [Mg/Fe] are correlated with
the Hubble type: they decrease for late type bulges. The predicted
spectral indices Mg$_{2}$ and $Fe52$, computed by evolutionary
synthesis models, are also similar to those observed.

\end{document}